\documentclass{article}
\usepackage{nips15submit_e}

\usepackage{graphicx}
\usepackage{subfigure}
\usepackage{color}

\usepackage{amsmath}
\usepackage{amsfonts}
\usepackage{xspace}

\usepackage{natbib}

\usepackage{hyperref}

\usepackage[utf8]{inputenc}

\newcommand{\ie}{\emph{i.e.}\xspace}
\newcommand{\eg}{\emph{e.g.}\xspace}

\title{Ethnicity sensitive author disambiguation using semi-supervised learning}

\author{Gilles Louppe\\
        CERN\\
        Switzerland\\
\And Hussein Al-Natsheh\\
        CERN\\
        Switzerland\\
\And Mateusz Susik\\
        CERN\\
        Switzerland\\
\And Eamonn Maguire\\
        CERN\\
        Switzerland}
\date{}

\nipsfinalcopy

\begin{document}

\maketitle

\begin{abstract}

Author name disambiguation in bibliographic databases is the problem of
grouping together scientific publications written by the same person, accounting
for potential homonyms and/or synonyms.
Among solutions to this problem, digital libraries are increasingly offering tools
for authors to manually curate their publications and claim those that are theirs.
Indirectly, these tools allow for the inexpensive collection of large annotated training
data, which can be further leveraged to build a complementary automated
disambiguation system capable of inferring patterns for identifying
publications written by the same person.
Building on more than 1 million publicly released crowdsourced annotations,
we propose an automated author disambiguation solution exploiting this data (i) to learn
an accurate classifier for identifying coreferring authors and (ii) to guide the clustering
of scientific publications by distinct authors in a semi-supervised way.
To the best of our knowledge, our analysis is the first to be carried out on data of
this size and coverage.
With respect to the state of the art, we validate the general pipeline used in most existing
solutions, and improve by: (i) proposing phonetic-based blocking strategies, thereby
increasing recall; and (ii) adding strong ethnicity-sensitive features for learning a linkage
function, thereby tailoring disambiguation to non-Western author names whenever necessary.

\end{abstract}


\section{Introduction}
\label{introduction}


In academic digital libraries, author name disambiguation is the problem of
grouping together publications written by the same person.
Author name disambiguation is often a difficult problem because an author may use different
spellings or name variants across their career (synonymy) and/or distinct authors may
share the same name (polysemy).
Most notably, author disambiguation is often more troublesome for researchers
from non-Western cultures, where personal names may be traditionally less diverse (leading to
homonym issues) or for which transliteration to Latin characters may not be unique (leading to
synonym issues).
With the fast growth of the scientific literature, author disambiguation has become a pressing
issue since the accuracy of information managed at the level of individuals directly affects:
the relevance search of results (\eg, when querying for all publications written by a given author);
the reliability of bibliometrics and author rankings (\eg, citation counts or other impact
metrics, as studied in \citep{strotmann2012author}); and/or the relevance of scientific network analysis
\citep{newman2001structure}.

Efforts and solutions to author disambiguation have been proposed from various
communities \citep{liu2014author}. On the one hand, libraries have maintained
authorship control through manual curation, either in a centralized way by
hiring professional collaborators or through developing services that invite authors to register
their publications themselves (\eg, Google Scholar or Inspire-HEP).
Recent efforts to create persistent digital identifiers assigned to researchers (\eg, ORCID or ResearcherID),
with the objective to embed these identifiers in the submission workflow of publishers
or repositories (\eg, Elsevier, arXiv or Inspire-HEP), would univocally solve any disambiguation issue.
With the large cost of centralized manual authorship control, or until crowdsourced solutions are more
widely adopted, the impact of these efforts are unfortunately limited by the efficiency, motivation
and integrity of their active contributors. Similarly, the success of persistent digital identifier efforts
is conditioned to a large and ubiquitous adoption by both researchers and publishers.
For these reasons, fully automated machine learning-based methods have been proposed during the past decade to
provide immediate, less costly, and satisfactory solutions to author
disambiguation. In this work, our goal is to explore and demonstrate how both
approaches can coexist and benefit from each other.
In particular, we study how labeled data obtained through manual curation (either centralized or
crowdsourced) can be exploited (i) to learn an accurate classifier for
identifying coreferring authors, and (ii) to guide the clustering of scientific
publications by distinct authors in a semi-supervised way.
Our analysis of parameters and features of this large dataset reveal that the general pipeline
commonly used in existing solutions is an effective approach for author disambiguation.
Additionally, we propose alternative strategies for blocking based on the
phonetization of author names to increase recall.
We also propose ethnicity-sensitive features for learning a linkage function,
thereby tailoring author disambiguation to non-Western author names whenever necessary.

The remainder of this report is structured as follows. In Section~\ref{related-works},
we briefly review machine learning solutions for author disambiguation.
The components of our method are then defined in Section~\ref{methods}
and its implementation described in Section~\ref{implementation}. Experiments
are carried out in Section~\ref{experiments}, where we explore and validate
features for the supervised learning of a linkage function and compare
strategies for the semi-supervised clustering of publications.
Finally, conclusions and future works are discussed in Section~\ref{conclusions}.


\section{Related work}
\label{related-works}

As reviewed in \citep{smalheiser2009author,ferreira2012brief,levin2012citation}, author
disambiguation algorithms are usually composed of two main components: (i) a
linkage function determining whether two publications have been written by the
same author; and (ii) a clustering algorithm producing clusters of publications
assumed to be written by the same author.
Approaches can be classified along several axes, depending on the type and
amount of data available, the way the linkage function is learned or defined, or the
clustering procedure used to group publications.
Methods relying on supervised learning usually make use of a small set of hand-labeled pairs
of publications identified as being either from the same or different authors to automatically learn a linkage
function between publications \citep{han2004two,huang2006efficient, culotta2007author,treeratpituk2009disambiguating,tran2014author}.

Training data is usually not easily available, therefore unsupervised approaches propose
the use of domain-specific, manually designed, linkage functions tailored towards author
disambiguation \citep{malin2005unsupervised,mcrae2006also,song2007efficient,
soler2007separating, kang2009co,fan2011graph,schulz2014exploiting}.
These approaches have the advantage of not requiring hand-labeled data, but generally do
not perform as well as supervised approaches.
To reconcile both worlds, semi-supervised methods make use of small, manually verified, clusters of
publications and/or high-precision domain-specific rules to build a training
set of pairs of publications, from which a linkage function is then built using supervised learning
\citep{ferreira2010effective,torvik2009author,levin2012citation}.

Semi-supervised approaches also allow for the tuning of the
clustering algorithm when the latter is applied to a mixed set of labeled
and unlabeled publications, \eg, by maximizing some clustering performance
metric on the known clusters \citep{levin2012citation}.

Due to the lack of large and publicly available datasets of curated
clusters of publications, studies on author disambiguation are usually
constrained to validating their results on manually built datasets of limited
size and scope (from a few hundred to a few thousand papers, with sparse
coverage of ambiguous cases), making the true performance of these methods
often difficult to assess with high confidence.
Additionally, despite devoted efforts to construct them, these datasets are rarely public,
making it even more difficult to compare methods using a common benchmark.

In this context, we position the work in this paper as a
semi-supervised solution for author disambiguation, with the significant
advantage of having a very large collection of more than 1 million crowdsourced annotations
of publications whose true authors are identified.
The extent and coverage of this data allows us to revisit, validate and nuance previous
findings regarding supervised learning of linkage functions, and to better explore strategies
for semi-supervised clustering.
Furthermore, by releasing our data in the public domain, we hope to provide a benchmark on
which further research on author disambiguation and related topics can be built.


\section{Semi-supervised author disambiguation}
\label{methods}

\begin{figure}
\centering
\includegraphics[width=\textwidth]{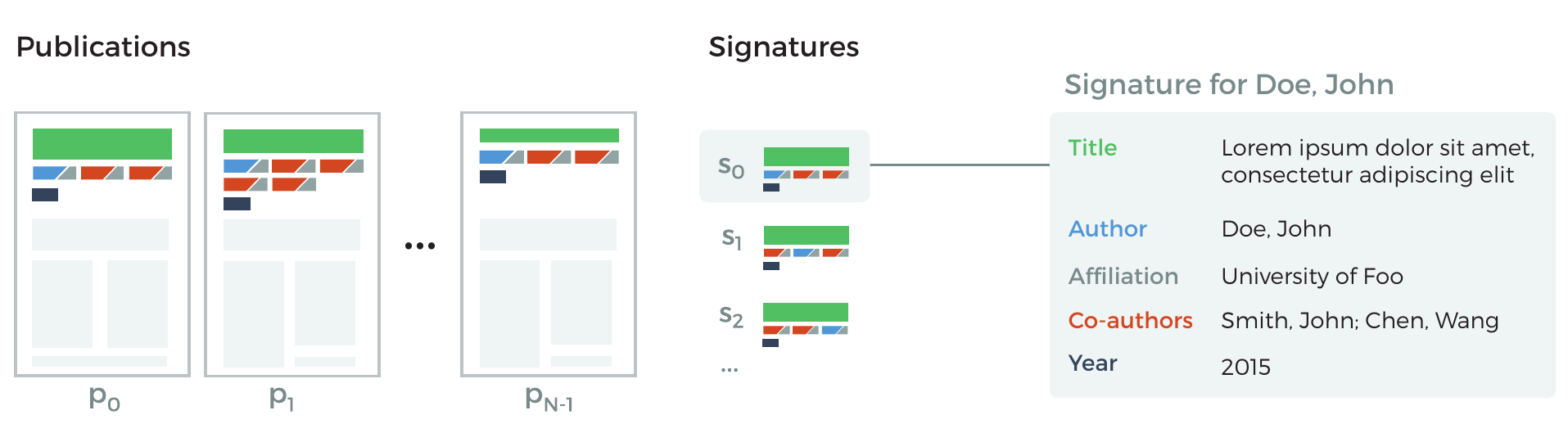}
\caption{An example signature $s$ for "Doe, John". A \textit{signature} is
defined as unique piece of information identifying an author on a publication,
along with any other metadata that can be derived from it, such as publication
title, co-authors or date of publication.}
\label{fig:signature}
\end{figure}

Formally, let us assume a set of publications ${\cal P} = \{ p_0, ...,
p_{N-1}\}$ along with the set of unique individuals ${\cal A} = \{ a_0, ...,
a_{M-1}\}$ having together authored all publications in ${\cal P}$.  Let us
define a signature $s \in p$ from a publication as a unique piece of
information identifying one of the authors of $p$ (\eg, the author name, his
affiliation, along with any other metadata that can be derived from $p$, as illustrated in Figure~\ref{fig:signature}). Let us
denote by ${\cal S} = \{ s | s \in p, p \in {\cal P} \}$ the set of all
signatures that can be extracted from all publications in ${\cal P}$.

In this
framework, author disambiguation can be stated as the problem of
finding a partition ${\cal C} = \{ c_0, ..., c_{M-1} \}$ of ${\cal S}$ such
that ${\cal S} = \cup_{i=0}^{M-1} c_i$, $c_i \cap c_j = \phi$ for all $i \neq
j$, and where subsets $c_i$, or clusters, each corresponds to the set of all
signatures belonging to the same individual $a_i$. Alternatively, the set
${\cal A}$ may remain (possibly partially) unknown, such that author
disambiguation boils down to finding a partition ${\cal C}$ where
subsets $c_i$  each correspond to the set of all signatures from the same
individual (without knowing who). Finally, in the case of partially annotated databases as studied in
this work, the set extends with the partial knowledge ${\cal C}^\prime = \{ c_0^\prime, ..., c_{M-1}^\prime \}$ of ${\cal C}$,
such that $c_i^\prime \subseteq c_i$, where $c_i^\prime$ may be empty.
Or put otherwise, the set extends with the assumption that all signatures
$s \in c_i^\prime$ belong to the same author.

\begin{figure}
\centering
\includegraphics[width=\textwidth]{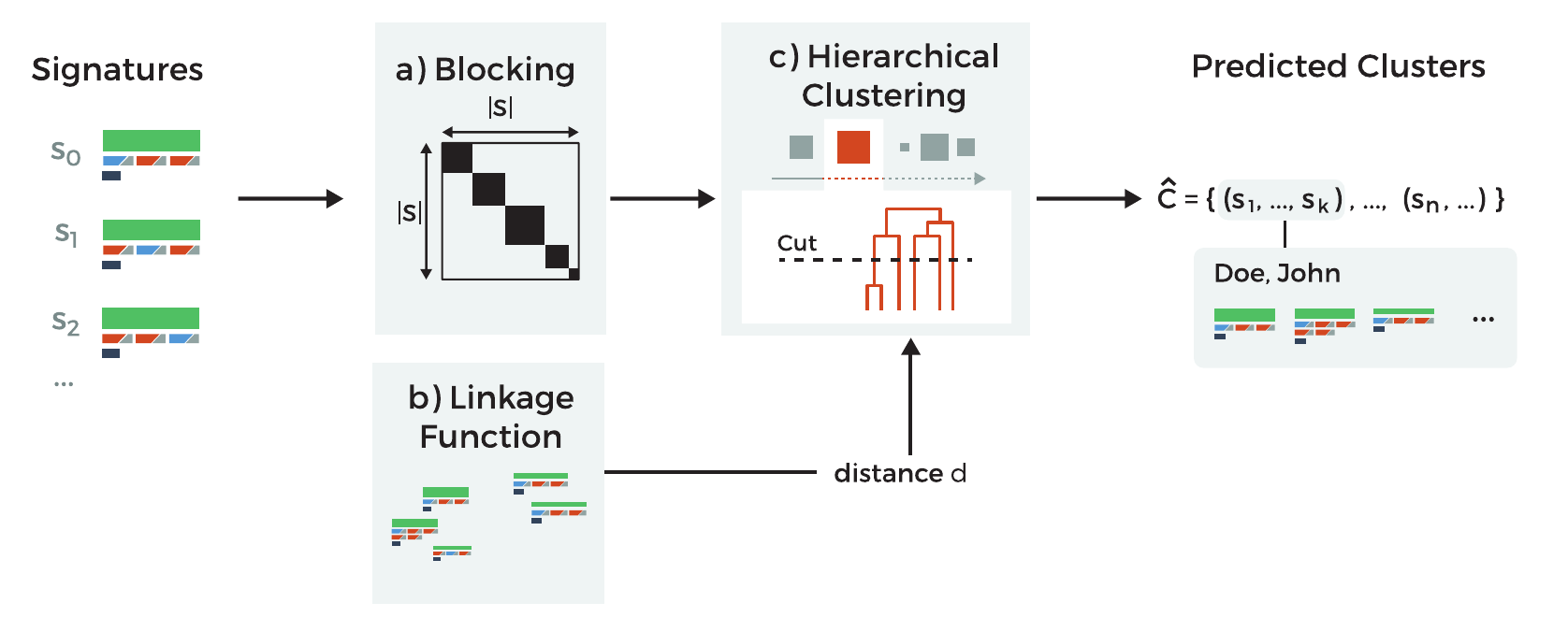}
\caption{Pipeline for author disambiguation: (a)
signatures are \textit{blocked} to reduce computational complexity, (b) a linkage
function is built with supervised learning, (c) independently within each block, signatures
are grouped using hierarchical agglomerative clustering.}
\label{fig:workflow}
\end{figure}

Inspired by several previous works described in Section~\ref{related-works},
we cast in this work author disambiguation into a semi-supervised clustering
problem.
Our algorithm is composed of three parts, as sketched in Figure~\ref{fig:workflow}: (i) a blocking
scheme whose goal is to roughly pre-cluster signatures ${\cal S}$ into smaller groups in order to
reduce computational complexity; (ii) the construction of a linkage function
$d$ between signatures using supervised learning; and (iii) the
semi-supervised clustering of all signatures within the same block, using $d$ as a pseudo distance metric.

\subsection{Blocking}
\label{methods:blocking}

As in previous works, the first part of our algorithm consists of dividing
signatures ${\cal S}$ into disjoint subsets ${\cal S}_{b_0}, ..., {\cal
S}_{b_{K-1}}$, or \textit{blocks} \citep{fellegi69}, followed by carrying out
author disambiguation on each one of these blocks independently.
By doing so, the computational complexity of clustering (see Section~\ref{methods:clustering})
typically reduces from $O(|{\cal S}|^2)$ to $O(\sum_b |{\cal S}_b|^2)$, which is much more
tractable as the number of signatures increases.
Since disambiguation is performed independently per block, a good blocking strategy should be
designed such that signatures from the same author are all mapped to the same
block, otherwise their correct clustering would not be possible in later stages of the workflow.
As a result, blocking should be a balance between reduced complexity and maximum recall.

The simplest and most common strategy for blocking, referred to hereon in as \textit{Surname and First Initial (SFI)},
groups signatures together if they share the same surname(s) and the same first
given name initial (\eg, \emph{SFI}("Doe, John") $==$ "Doe, J").
Despite satisfactory performance, there are several cases where this simple strategy fails to cluster related pairs of signatures together, including:

\begin{enumerate}
  \item There are different
  ways of writing an author name, or signatures contain a typo
  (\eg, "Mueller, R." and "Muller, R.", "Tchaikovsky, P." and "Czajkowski, P.").

  \item An author has multiple surnames and some signatures place the first part of the surname within the given names (\eg, "Martinez Torres, A." and "Torres, A. Martinez").

  \item An author has multiple surnames and, on some signatures, only the first surname is
  present (\eg, "Smith-Jones, A." and "Smith, A.")

  \item An author has multiple given names and they are not always all recorded (\eg,
  "Smith, Jack" and "Smith, A. J.")

  \item An authors surname changed (\eg, due to marriage).
\end{enumerate}

To account for these issues we propose instead to block signatures based on the
phonetic representation of the normalized surname.
Normalization involves stripping accents (\eg, "Jabłoński, Ł" $\rightarrow$ "Jablonski, L") and name
affixes that inconsistently appear in signatures (\eg, "van der Waals, J. D."
$\rightarrow$ "Waals, J. D."), while phonetization is based either on the
Double Metaphone \citep{doublemetaphone}, the NYSIIS \citep{nysiis} or the
Soundex \citep{Soundex} phonetic algorithms for mapping author names to their pronunciations.
Together, these processing steps allow for grouping of most name variants of the same
person in the same block with a small increase in the overall computational complexity, thereby solving case 1.

In the case of multiple surnames (cases 2 and 3), we propose to block
signatures in two phases.
In the first phase, all the signatures with a single surname are clustered together.
Every different surname token creates a new block.
In the second phase, the signatures with multiple surnames are compared
with the blocks for the first and last surname.
If the first surnames of an author were already used as the last given names on some of the signatures, the
new signature is assigned to the block of the last surname (case 2).
Otherwise, the signature is assigned to the block of the first surname (case 3).
Finally, to prevent the creation of too large blocks, signatures are further divided
along their first given name initial.
Cases 4 and 5 are not explicitly handled.

\subsection{Linkage function}
\label{methods:linkage}

\textit{Supervised classification.} The second part of the algorithm is the
automatic construction of a pair-wise linkage function between signatures for use
during the clustering step which groups all signatures from the same author.

Formally, the goal is to build a function $d: {\cal S} \times {\cal S} \mapsto
[0, 1]$, such that $d(s_1, s_2)$ approaches $0$ if both signatures $s_1$ and
$s_2$ belong to the same author, and $1$ otherwise.
This problem can be cast as a standard supervised classification task, where inputs
are pairs of signatures and outputs are classes $0$ (same authors), and $1$
(distinct authors). In this work, we evaluate Random Forests (RF, \cite{breiman2001random}),
Gradient Boosted Regression Trees (GBRT, \cite{friedman2001greedy}),
and Logistic Regression \citep{fan2008liblinear} as classifiers.

\textit{Input features.} In most cases, supervised learning algorithms assume
the input space ${\cal X}$ to be numeric (\eg, $\mathbb{R}^p$), making them
not directly applicable to structured input spaces such as ${\cal S} \times
{\cal S}$.
Following previous works, pairs of signatures $(s_1, s_2)$ are first transformed to vectors $v \in \mathbb{R}^p$
by building so-called similarity profiles \citep{treeratpituk2009disambiguating} on which supervised learning is carried out.
In this work, we design and evaluate fifteen standard input
features based on the comparison of signature fields, as reported in the first
half of Table~\ref{table:features}.
As an illustrative example, the \textit{Full name} feature corresponds to the similarity between the (full)
author name fields of the two signatures, as measured using as combination
operator the cosine similarity between their respective $(n,m)$-\emph{TF-IDF} vector
representations\footnote{$(n,m)$ denotes that the \emph{TF-IDF} vectors are computed
from character $n$, $n+1$, ..., $m$-grams.
When not specified, \emph{TF-IDF} vectors are otherwise computed from words.}.
Similarly, the \textit{Year difference} feature measures the absolute difference between the publication date of the
articles to which the two signatures respectively belong.

Author names from different cultures, origins or ethnic groups are likely to be
disambiguated using different strategies (\eg, pairs of signatures with French
author names versus pairs of signatures with Chinese author names) \citep{treeratpituk2012name, chin2014effective}.
To support our disambiguation algorithm, we added seven features to our feature set, with each
evaluating the degree of belonging of both signatures to an ethnic group,
as reported in the second half of Table~\ref{table:features}.

More specifically, using census data extracted from \citep{rugglesintegrated},
we build a support vector machine classifier (using a linear kernel and
one-versus-all classification scheme) for mapping the
 $(1,5)$-TF-IDF representation of an author name to one of the seven ethnic groups. Given a
pair of signatures $(s_1, s_2)$, the proposed ethnicity features are each
computed as the estimated probability of $s_1$ belonging to the corresponding
ethnic group, multiplied by the estimated probability of $s_2$ belonging to the
same group. In doing so, the expectation is for the linkage function to become
sensitive to the actual origin of the authors depending on the values of these
features. Indirectly, let us also note that these features hold discriminative
power since if author names are predicted to belong to different ethnic groups,
then they are also likely to correspond to distinct people.

\begin{table}
\caption{Input features for learning a linkage function}
\label{table:features}
\centering
\begin{tabular}{|l|l|}
  \hline
  \textbf{Feature} & \textbf{Combination operator}\\
  \hline
  \hline
  Full name & Cosine similarity of $(2,4)$-TF-IDF\\
  Given names & Cosine similarity of $(2,4)$-TF-IDF\\
  First given name & Jaro-Winkler distance\\
  Second given name & Jaro-Winkler distance\\
  Given name initial & Equality\\
  Affiliation & Cosine similarity of $(2,4)$-TF-IDF\\
  Co-authors & Cosine similarity of TF-IDF\\
  Title & Cosine similarity of $(2,4)$-TF-IDF\\
  Journal & Cosine similarity of $(2,4)$-TF-IDF\\
  Abstract & Cosine similarity of TF-IDF\\
  Keywords & Cosine similarity of TF-IDF\\
  Collaborations & Cosine similarity of TF-IDF\\
  References & Cosine similarity of TF-IDF\\
  Subject & Cosine similarity of TF-IDF\\
  Year difference & Absolute difference\\
  \hline
  White & Product of estimated probabilities\\
  Black & Product of estimated probabilities\\
  American Indian or Alaska Native & Product of estimated probabilities\\
  Chinese & Product of estimated probabilities\\
  Japanese & Product of estimated probabilities\\
  Other Asian or Pacific Islander & Product of estimated probabilities\\
  Others & Product of estimated probabilities\\
  \hline
\end{tabular}
\end{table}


\textit{Building a training set.} The distinctive aspect of our work is the
knowledge of more than 1 million crowdsourced annotations ${\cal C}^\prime = \{
c_0^\prime, ..., c_{M-1}^\prime \}$, indicating together that all signature $s \in
c_i^\prime$ are known to correspond to the same individual $a_i$.
In particular, this data can be used to generate positive pairs $(x=(s_1, s_2), y=0)$ for all
$s_1, s_2 \in c_i^\prime$, for all $i$. Similarly, negative pairs $(x=(s_1,
s_2), y=1)$ can be extracted for all $s_1 \in c_i^\prime, s_2 \in c_j^\prime$, for
all $i \neq j$.

The most straightforward approach for building a training set on which to learn
a linkage function is to sample an equal number of positive and negative pairs,
as suggested above.
By observing that the linkage function $d$ will eventually
be used only on pairs of signatures from the same block $S_b$, a further
refinement for building a training set is to restrict positive and negative
pairs $(s_1, s_2)$ to only those for which $s_1$ and $s_2$ belong to the same
block. In doing so, the trained classifier is forced to learn intra-block
discriminative patterns rather than inter-block differences.
Furthermore, as noted in \citep{lange2011frequency}, most signature pairs are non-ambiguous:
if both signatures share the same author names, then
they correspond to the same individual, otherwise they do not.
Rather than sampling pairs uniformly at random, we propose to oversample
difficult cases when building the training set (\ie, pairs of signatures with
different author names corresponding to same individual, and pairs of
signatures with identical author names but corresponding to distinct
individuals) in order to improve the overall accuracy of the linkage function.

\subsection{Semi-supervised clustering}
\label{methods:clustering}

The last component of our author disambiguation pipeline is clustering, that is
the process of grouping together, within a block, all signatures from the same
individual (and only those).
As for many other works on author disambiguation, we make use of hierarchical clustering \citep{ward1963hierarchical} for
building clusters of signatures in a bottom-up fashion.
The method involves iteratively merging together the two most similar clusters until all clusters
are merged together at the top of the hierarchy.
Similarity between clusters is evaluated using either complete, single or average linkage, using
as a pseudo-distance metric the probability that $s_1$ and $s_2$ correspond to
distinct authors, as calculated from the custom linkage function $d$ from Section \ref{methods:linkage}.

\begin{figure}
\centering
\includegraphics[width=\textwidth]{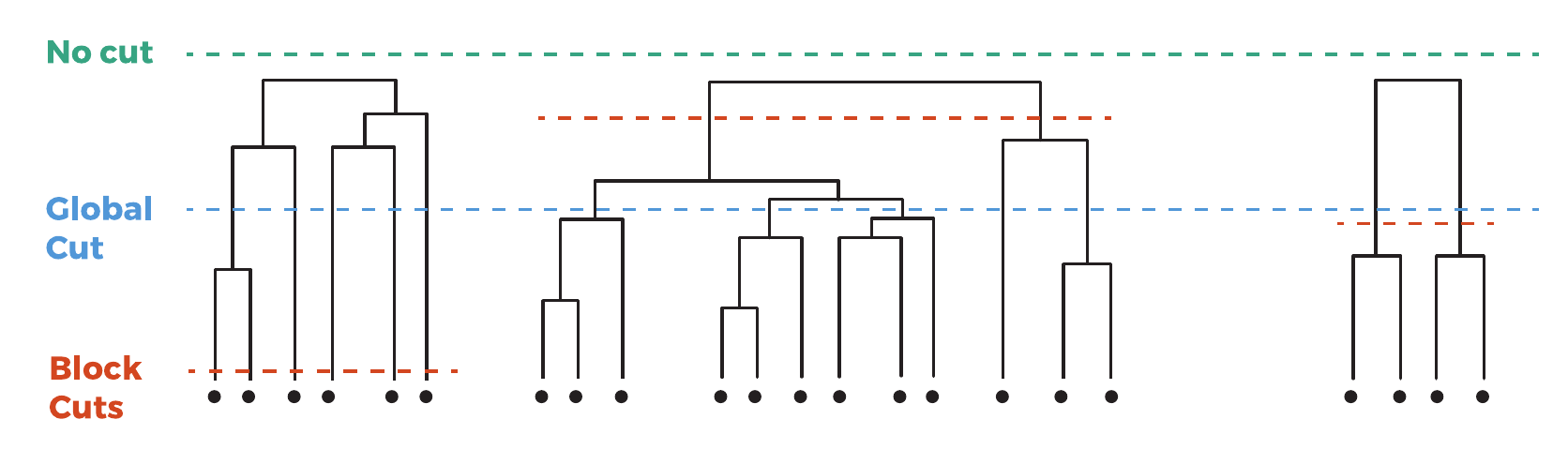}
\caption{Semi-supervised cut-off strategies to form flat clusters of signatures.}
\label{fig:cuts}
\end{figure}

To form flat clusters from the hierarchy, one must decide on a maximum
distance threshold above which clusters are considered to correspond to
distinct authors.
Let us denote by ${\cal S}^\prime = \{ s | s \in c^\prime, c^\prime
\in {\cal C}^\prime \}$ the set of all signatures for which partial clusters are
known.
Let us also denote by $\smash{\widehat{\cal C}}$ the predicted clusters for all signatures in ${\cal S}$, and by
$\smash{\widehat{\cal C}^\prime} = \{ \widehat{c} \cap {\cal S}^\prime | \widehat{c} \in \widehat{\cal C} \}$
the predicted clusters restricted to signatures for which partial clusters are known.
From these, we evaluate the following semi-supervised cut-off strategies, as illustrated in Figure~\ref{fig:cuts}:
\begin{itemize}

\item \textit{No cut:} all signatures from the same block are assumed to be from the same author.

\item \textit{Global cut:} the threshold is chosen globally over all blocks,
    as the one maximizing some score $f({\cal C}^\prime, \widehat{\cal C}^\prime)$.

\item \textit{Block cut:} the threshold is chosen locally at each block $b$,
    as the one maximizing some score $f({\cal C}_b^\prime, \widehat{\cal C}_b^\prime)$.
    In case ${\cal C}_b^\prime$ is empty, then all signatures from $b$ are clustered together.
\end{itemize}

\section{Implementation}
\label{implementation}

As part of this work, we developed a stand-alone application for author
disambiguation, publicly available online\footnote{\url{https://github.com/glouppe/beard}} for free reuse
or study.
Our implementation builds upon the Python scientific stack, making
use of the Scikit-Learn library \citep{scikitlearn} for the supervised learning
of a linkage function and of SciPy \citep{scipy} for clustering.
All components of the disambiguation pipeline have been designed to follow the
Scikit-Learn API \citep{scikitlearnAPI}, making them easy to maintain,
understand and reuse.
Our implementation is made to be efficient, exploiting parallelization when available, and ready for production environments.
It is also designed to be runnable
in an incremental fashion, by enabling disambiguation only on specified blocks if desired,
instead of having to run the disambiguation process on the whole signature set.


\section{Experiments}
\label{experiments}

\subsection{Data}

The author disambiguation solution proposed in this work, along with its
enhancements, are evaluated on data extracted from the \emph{INSPIRE} portal
\citep{gentil2009information}, a digital library for scientific literature in
high-energy physics.
Overall, the portal holds more than 1 million publications ${\cal P}$,
forming in total a set ${\cal S}$ of more than 10 million signatures.
Out of these, around 13\% have been \textit{claimed} by their
original authors, marked as such by professional curators or automatically assigned to their true authors thanks
to persistent identifiers provided by publishers or other sources.
Together, they constitute a trusted set $({\cal S}^\prime, {\cal C}^\prime)$ of 15388 distinct individuals sharing
36340 unique author names spread within 1201763 signatures on 360066
publications. This data covers several decades in time and dozens of author
nationalities worldwide.

Following the \emph{INSPIRE} terms of use, the signatures ${\cal S}^\prime$ and their
corresponding clusters ${\cal C}^\prime$ are released
online\footnote{\url{https://github.com/glouppe/paper-author-disambiguation}}
under the CC0 license.
To the best of our knowledge, data of this size and coverage is the first to be publicly
released in the scope of author disambiguation research.

\subsection{Evaluation protocol}

Experiments carried out to study the impact of the proposed algorithmic
components and refinements, as described in Section~\ref{methods}, follow a
standard 3-fold cross-validation protocol, using $({\cal S}^\prime, {\cal
C}^\prime)$ as ground-truth dataset. To replicate the $|{\cal S}^\prime| /
|{\cal S}| \approx 13\%$ ratio of claimed signatures with respect to the total
set of signatures, as on the INSPIRE platform, cross-validation folds are
constructed by sampling 13\% of claimed signatures to form a training set ${\cal
S}_\text{train}^\prime \subseteq {\cal S}^\prime$.
The remaining signatures ${\cal S}_\text{test}^\prime = {\cal S}^\prime \setminus {\cal
S}_\text{train}^\prime$ are used for testing.
Therefore, ${\cal C}_\text{train}^\prime = \{ c^\prime \cap {\cal S}_\text{train}^\prime | c^\prime \in {\cal C}^\prime
\}$ represents the partial known clusters on the training fold, while ${\cal
C}_\text{test}^\prime$ are those used for testing.

As commonly performed in author disambiguation research,
we evaluate the predicted clusters over testing data  ${\cal C}_\text{test}^\prime$,
using both B3 and pairwise precision, recall
and F-measure, as defined below:
\begin{align}
P_\text{B3}({\cal C}, \widehat{\cal C}, {\cal S}) &= \frac{1}{|{\cal S}|} \sum_{s \in {\cal S}} \frac{|c(s) \cap \widehat{c}(s)|}{|\widehat{c}(s)|} \\
R_\text{B3}({\cal C}, \widehat{\cal C}, {\cal S}) &= \frac{1}{|{\cal S}|} \sum_{s \in {\cal S}} \frac{|c(s) \cap \widehat{c}(s)|}{|c(s)|}\\
F_\text{B3}({\cal C}, \widehat{\cal C}, {\cal S}) &= \frac{2 P_\text{B3}({\cal C}, \widehat{\cal C}, {\cal S}) R_\text{B3}({\cal C}, \widehat{\cal C}, {\cal S})}{P_\text{B3}({\cal C}, \widehat{\cal C}, {\cal S}) + P_\text{B3}({\cal C}, \widehat{\cal C}, {\cal S})}\\
P_\text{pairwise}({\cal C}, \widehat{\cal C}) &= \frac{|p({\cal C}) \cap p(\widehat{\cal C})|}{|p(\widehat{\cal C})|}\\
R_\text{pairwise}({\cal C}, \widehat{\cal C}) &= \frac{|p({\cal C}) \cap p(\widehat{\cal C})|}{|p({\cal C})|}\\
F_\text{pairwise}({\cal C}, \widehat{\cal C}) &= \frac{2 P_\text{pairwise}({\cal C}, \widehat{\cal C}) R_\text{pairwise}({\cal C}, \widehat{\cal C})}{P_\text{pairwise}({\cal C}, \widehat{\cal C}) + R_\text{pairwise}({\cal C}, \widehat{\cal C})}
\end{align}
and where $c(s)$ (resp. $\widehat{c}(s)$) is the cluster $c \in {\cal C}$ such that
$s \in c$ (resp. the cluster $\widehat{c} \in \widehat{\cal C}$ such that $s
\in \widehat{c}$), and where $p({\cal C}) = \cup_{c \in {\cal C}} \{ (s_1, s_2)
| s_1, s_2 \in c, s_1 \neq s_2 \}$ is the set of all pairs of signatures from
the same clusters in ${\cal C}$.
Precision evaluates whether signatures are grouped only with signatures from the same true clusters,
while recall measures the extent to which all signatures from the same true clusters are
effectively grouped together.
The F-measure is the harmonic mean between these two quantities.
In the analysis below, we rely primarily on the B3 F-measure for discussing results, as the pairwise variant
tends to favor large clusters (because the number of pairs is quadratic with the cluster size),
hence unfairly giving preference to authors with many publications.
By contrast, the B3 F-measure weights clusters linearly with respect to their size.
General conclusions drawn below remain however consistent for pairwise F.

\subsection{Results and discussion}

\begin{table}
\caption{Average precision, recall and f-measure scores on test folds.}
\label{table:results}
\centering
\begin{tabular}{|l|c c c|c c c|}
  \hline
                       & \multicolumn{3}{|c|}{\textbf{B3}} & \multicolumn{3}{|c|}{\textbf{Pairwise}} \\
  \textbf{Description} & $P$ & $R$ & $F$ & $P$ & $R$ & $F$\\
  \hline
  \hline
Baseline & 0.9901 & 0.9760 & 0.9830 & 0.9948 & 0.9738 & 0.9842 \\
\hline
Blocking = SFI & 0.9901 & 0.9760 & 0.9830 & 0.9948 & 0.9738 & 0.9842 \\
Blocking = Double metaphone & 0.9856 & 0.9827 & 0.9841 & 0.9927 & 0.9817 & 0.9871 \\
Blocking = NYSIIS & 0.9875 & 0.9826 & \textbf{0.9850} & 0.9936 & 0.9814 & \textbf{0.9875} \\
Blocking = Soundex & 0.9886 & 0.9745 & 0.9815 & 0.9935 & 0.9725 & 0.9828 \\
\hline
No name normalization & 0.9887 & 0.9697 & 0.9791 & 0.9931 & 0.9658 & 0.9793 \\
Name normalization & 0.9901 & 0.9760 & \textbf{0.9830} & 0.9948 & 0.9738 & \textbf{0.9842} \\
\hline
Classifier = GBRT & 0.9901 & 0.9760 & 0.9830 & 0.9948 & 0.9738 & 0.9842 \\
Classifier = Random Forests & 0.9909 & 0.9783 & \textbf{0.9846} & 0.9957 & 0.9752 & \textbf{0.9854} \\
Classifier = Linear Regression & 0.9749 & 0.9584 & 0.9666 & 0.9717 & 0.9569 & 0.9643 \\
\hline
Training pairs = Non-blocked, uniform & 0.9793 & 0.9630 & 0.9711 & 0.9756 & 0.9629 & 0.9692 \\
Training pairs = Blocked, uniform & 0.9854 & 0.9720 & 0.9786 & 0.9850 & 0.9707 & 0.9778 \\
Training pairs = Blocked, balanced & 0.9901 & 0.9760 & \textbf{0.9830} & 0.9948 & 0.9738 & \textbf{0.9842} \\
\hline
Clustering = Average linkage & 0.9901 & 0.9760 & \textbf{0.9830} & 0.9948 & 0.9738 & \textbf{0.9842} \\
Clustering = Single linkage & 0.9741 & 0.9603 & 0.9671 & 0.9543 & 0.9626 & 0.9584 \\
Clustering = Complete linkage & 0.9862 & 0.9709 & 0.9785 & 0.9920 & 0.9688 & 0.9803 \\
\hline
No cut & 0.9024 & 0.9828 & 0.9409 & 0.8298 & 0.9776 & 0.8977 \\
Global cut & 0.9892 & 0.9737 & 0.9814 & 0.9940 & 0.9727 & 0.9832 \\
Block cut & 0.9901 & 0.9760 & \textbf{0.9830} & 0.9948 & 0.9738 & \textbf{0.9842} \\
\hline
Combined best settings & 0.9888 & 0.9848 & \textbf{0.9868} & 0.9951 & 0.9831 & \textbf{0.9890} \\
  \hline
\end{tabular}
\end{table}

\textit{Baseline.} Results presented in Table~\ref{table:results} are discussed with
respect to a baseline solution using the following combination of components:
\begin{itemize}
\item Blocking: same surname and the same first given name initial strategy (SFI);
\item Linkage function: all 22 features defined in Table~\ref{table:features},
    gradient boosted regression trees as supervised learning algorithm
    and a training set of pairs built from $({\cal S}_\text{train}^\prime, {\cal C}_\text{train}^\prime)$, by balancing easy and difficult cases.
\item Clustering: agglomerative clustering using average linkage and
    block cuts found to maximize $F_\text{B3}({\cal C}_\text{train}^\prime, \widehat{\cal C}_\text{train}^\prime, {\cal S}_\text{train}^\prime)$.
\end{itemize}

\textit{Blocking.} The good precision of the baseline ($0.9901$), but its
lower recall ($0.9760$) suggest that the blocking strategy might be the
limiting factor to further overall improvements.
As shown in Table~\ref{table:blocking}, the maximum recall (\ie, if within a block, all signatures were clustered optimally) for SFI is $0.9828$.
At the price of fewer and therefore slightly larger blocks (as reported in the right column of Table~\ref{table:blocking}), the
proposed phonetic-based blocking strategies show better maximum recall (all
around $0.9905$), thereby pushing further the upper bound on the maximum
performance of author disambiguation.
Let us remind however that the reported maximum recalls for the blocking strategies using phonetization are
also raised due to the better handling of multiple surnames, as described in Section \ref{methods:blocking}.

As Table~\ref{table:results} shows, switching to either Double metaphone or NYSIIS phonetic-based blocking allows
to improve the overall F-measure score, trading precision for recall.
In particular, the NYSIIS-based phonetic blocking shows to be the most effective when applied
to the baseline (with an F-measure of $0.9850$) while also being the most
efficient computationally (with 10857 blocks versus 12978 for the baseline).

Finally, let us also note that Table~\ref{table:blocking} corroborates the
estimation of \citep{torvik2009author}, stating that SFI blocking has a recall
around $98\%$ on real data.

\begin{table}
\caption{Maximum recall $R_\text{B3}^*$ and $R_\text{pairwise}^*$ of blocking strategies, and their number of blocks on ${\cal S}^\prime$.}
\label{table:blocking}
\centering
\begin{tabular}{|l|cc|c|}
  \hline
  \textbf{Blocking} & $R_\text{B3}^*$ & $R_\text{pairwise}^*$ & \# blocks \\
  \hline
  \hline
    SFI & 0.9828 & 0.9776 & 12978 \\
    Double metaphone & 0.9907 & 0.9863 & 9753 \\
    NYSIIS & 0.9902 & 0.9861 & 10857 \\
    Soundex & 0.9906 & 0.9863 & 9403 \\
  \hline
\end{tabular}
\end{table}

\textit{Name normalization.} As discussed previously, the seemingly insignificant step of
normalizing author names (stripping accents, removing affixes), as performed in the
baseline, is shown to be important. Results from Table~\ref{table:results} clearly suggest that not
normalizing significantly reduces performance (yielding an F-measure of $0.9830$ when normalizing,
but decreasing to $0.9791$ when raw author name strings are used instead).

\textit{Linkage function.} Let us first comment on the results regarding the
supervised algorithm used to learn the linkage function.
As Table~\ref{table:results} indicates, both tree-based algorithms appear to be
significantly better fit than Linear Regression ($0.9830$ and $0.9846$ for GBRT
and Random Forests versus $0.9666$ for Linear Regression). This result is
consistent with \citep{treeratpituk2009disambiguating} which evaluated the use of
Random Forests for author disambiguation, but contradicts results of
\citep{levin2012citation} for which Logistic Regression appeared to be the best
classifier.
Provided hyper-parameters are properly tuned, the superiority of
tree-based methods is in our opinion not surprising.
Indeed, given the fact that the optimal linkage function is likely to be non-linear, non-parametric
methods are expected to yield better results, as the experiments here confirm.

Second, properly constructing a training set of positive and negative pairs of
signatures from which to learn a linkage function yields a significant
improvement.
A random sampling of positive and negative pairs, without taking
blocking into account, significantly impacts the overall performance
($0.9711$). When pairs are drawn only from blocks, performance increases
($0.9786$), which confirms our intuition that $d$ should be built only from
pairs it will be used to eventually cluster. Finally, making the classification
problem more difficult by oversampling complex cases proves to be relevant,
by further improving the disambiguation results ($0.9830$).

\begin{figure}
\centering
\caption{Recursive Feature Elimination analysis. }
\label{fig:rfe}
\includegraphics[width=\textwidth]{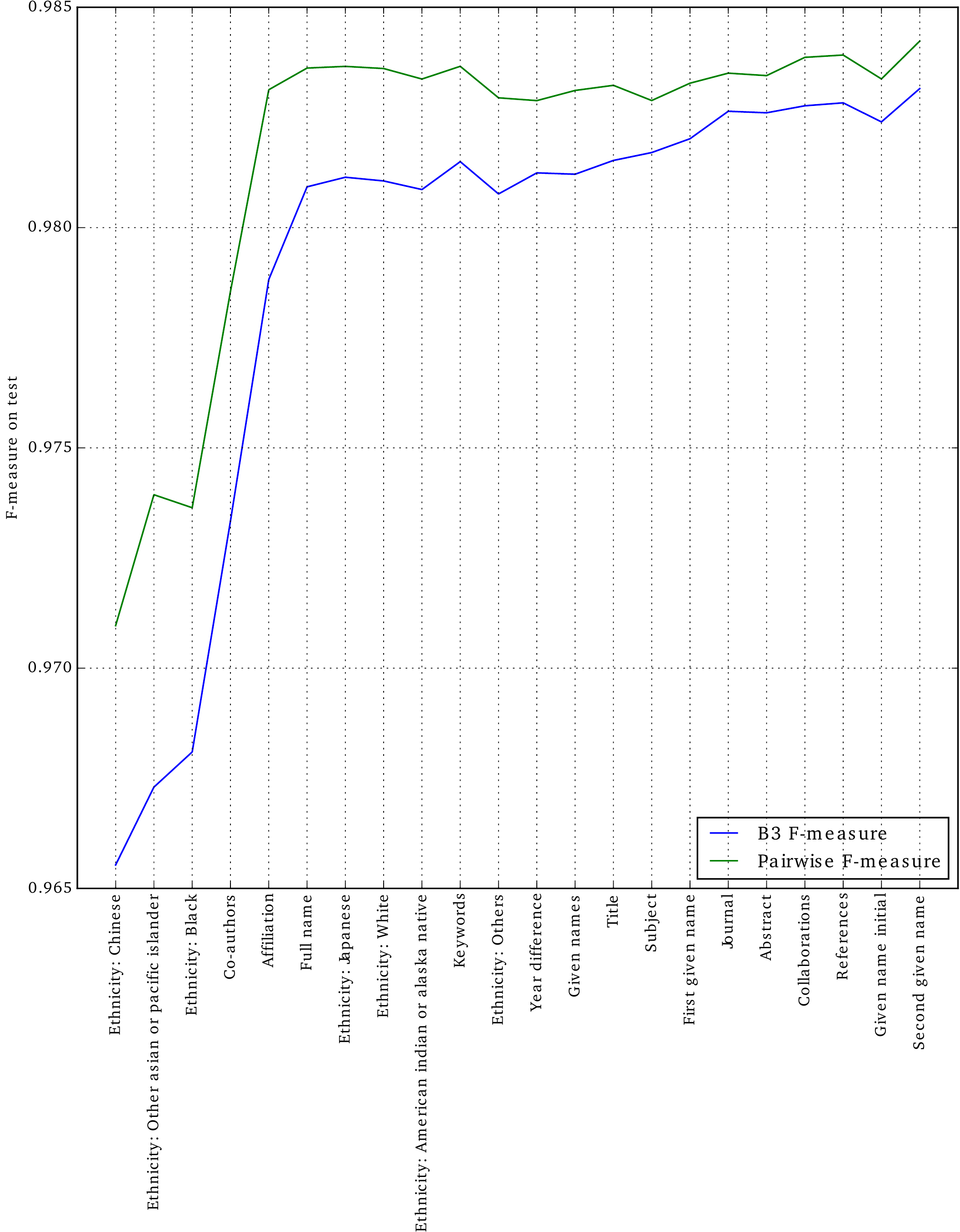}
\end{figure}

Using Recursive Feature Elimination \cite{guyon2002gene}, we next evaluate the
usefulness of all fifteen standard and seven additional ethnicity features for learning
the linkage function. The analysis consists in using the baseline algorithm
first using all twenty two features, to determine the least discriminative from feature
importances \citep{louppe2013understanding}, and then re-learn the baseline
using all but that one feature. That process is repeated recursively until
eventually only one feature remains. Results are presented in
Figure~\ref{fig:rfe} for one of the three folds, starting from the far right with
the baseline and \textit{Second given name} being the least important feature,
and ending on the left with all features eliminated but \textit{Chinese}. As
the figure illustrates, the most important features are ethnic-based features
(\textit{Chinese}, \textit{Other Asian}, \textit{Black}) along with
\textit{Co-authors}, \textit{Affiliation} and \textit{Full name}. Adding the remaining
other features only brings marginal improvements, with \textit{Journal},
\textit{Abstract}, \textit{Collaborations}, \textit{References}, \textit{Given
name initial} and \textit{Second given name} being almost insignificant.
Overall, these results highlight the added value of the proposed ethnicity
features.
Their duality in modeling both the similarity between author names
and their origins make them very strong predictors for author disambiguation.
The results also corroborate those from \citep{kang2009co} or \citep{ferreira2010effective}, who
found that the similarity between co-authors was a highly discriminative
feature.
If computational complexity is a concern, this analysis also
shows how decent performance can be achieved using only a very
small set of features, as also observed in
\citep{treeratpituk2009disambiguating} or \citep{levin2012citation}.

\textit{Semi-supervised clustering.} The last part of our experiment concerns
the study of agglomerative clustering and the best way to find a cut-off
threshold to form clusters. Results from Table~\ref{table:results}
first clearly indicate that average linkage is significantly better than
both single and complete linkage.

Clustering together all signatures from the same block is the least effective
strategy ($0.9409$), but yields anyhow surprisingly decent accuracy, given the
fact it requires almost no computation (\ie, both learning a linkage function
and running agglomerative clustering can be skipped -- only the blocking
function is needed to group signatures). In particular, this result reveals
that author names are not ambiguous in most cases\footnote{This holds for the
data we extracted, but may in the future, with the rise of non-Western
researchers, be an underestimate of the ambiguous cases.} and that only a small
fraction of them requires advanced disambiguation procedures. On the other
hand, both global and block cut thresholding strategies give very good results,
with a slight advantage for block cuts ($0.9814$ versus $0.9830$), as expected.
In case ${\cal S}^\prime_b$ is empty (\eg, because it corresponds to a young
researcher at the beginning of his career), this therefore suggests that either using
a cut-off threshold learned globally from the known data or using SFI would in general give
satisfactory results, only marginally worse than if claimed signatures had been
known.

\textit{Combined best settings.} When all best settings are combined (\ie,
Blocking = NYSIIS, Name normalization, Classifier = Random Forests, Training
pairs = blocked and balanced, Clustering = Average linkage, Block cuts),
performance reaches $0.9862$, \ie, the best of all reported results. In particular,
this combination exhibits both the high recall of phonetic blocking based on the NYSIIS algorithm
and the high precision of Random Forests.


\section{Conclusions}
\label{conclusions}

In this work, we have revisited and validated the general author disambiguation
pipeline introduced in previous independent research work.
The generic approach is composed of three components, whose design and tuning are all critical
to good performance: (i) a blocking function for pre-clustering signatures
and reducing computational complexity, (ii) a linkage function for identifying
signatures with coreferring authors and (iii) the agglomerative clustering of
signatures. Making use of a distinctively large dataset of more than 1 million
crowdsourced annotations, we experimentally study all three components and
propose further improvements. With regards to blocking, we suggest to use
phonetization of author names to increase recall while maintaining low
computational complexity. For the linkage function, we introduce
ethnicity-sensitive features for the automatic tailoring of disambiguation to non-Western
author names whenever necessary. Finally, we explore semi-supervised cut-off
threshold strategies for agglomerative clustering. For all three components,
experiments show that our refinements all yield significantly better author
disambiguation accuracy.

Overall, these results all encourage further improvements and research. For
blocking, one of the open challenges is to manage signatures with inconsistent
surnames or inconsistent first given names (cases 4 and 5, as described in
Section~\ref{methods:blocking}) while maintaining blocks to a tractable size.
As phonetic algorithms are not yet perfect, another direction  for further work is the design of better
phonetization functions, tailored for author disambiguation. For the linkage function,
the good results of the proposed features pave the way for further research  in
ethnicity-sensitive author disambiguation. The automatic fitting of the
pipeline to cultures and ethnic groups for which standard author disambiguation
is known to be less efficient (\eg, Chinese authors with many homonyms)  indeed
constitutes a direction of research with great potential benefits for the
concerned scientific communities.

As part of this study, we also publicly release the annotated data extracted
from the \emph{INSPIRE} platform, on which our experiments are based.
To the best of our knowledge, data of this size and coverage is the first to be
available in author disambiguation research. By releasing the data publicly,
we hope to provide the basis for further research on author disambiguation
and related topics.




\bibliographystyle{apalike}
\bibliography{bib}

\begin{thebibliography}{}

\bibitem[Breiman, 2001]{breiman2001random}
Breiman, L. (2001).
\newblock Random forests.
\newblock {\em Machine learning}, 45(1):5--32.

\bibitem[Buitinck et~al., 2013]{scikitlearnAPI}
Buitinck, L., Louppe, G., Blondel, M., Pedregosa, F., Mueller, A., Grisel, O.,
  Niculae, V., Prettenhofer, P., Gramfort, A., Grobler, J., Layton, R.,
  VanderPlas, J., Joly, A., Holt, B., and Varoquaux, G. (2013).
\newblock {API} design for machine learning software: experiences from the
  scikit-learn project.
\newblock {\em CoRR}, abs/1309.0238.

\bibitem[Chin et~al., 2014]{chin2014effective}
Chin, W.-S., Zhuang, Y., Juan, Y.-C., Wu, F., Tung, H.-Y., Yu, T., Wang, J.-P.,
  Chang, C.-X., Yang, C.-P., Chang, W.-C., et~al. (2014).
\newblock Effective string processing and matching for author disambiguation.
\newblock {\em The Journal of Machine Learning Research}, 15(1):3037--3064.

\bibitem[Culotta et~al., 2007]{culotta2007author}
Culotta, A., Kanani, P., Hall, R., Wick, M., and McCallum, A. (2007).
\newblock Author disambiguation using error-driven machine learning with a
  ranking loss function.
\newblock In {\em Sixth International Workshop on Information Integration on
  the Web (IIWeb-07), Vancouver, Canada}.

\bibitem[Fan et~al., 2008]{fan2008liblinear}
Fan, R.-E., Chang, K.-W., Hsieh, C.-J., Wang, X.-R., and Lin, C.-J. (2008).
\newblock Liblinear: A library for large linear classification.
\newblock {\em The Journal of Machine Learning Research}, 9:1871--1874.

\bibitem[Fan et~al., 2011]{fan2011graph}
Fan, X., Wang, J., Pu, X., Zhou, L., and Lv, B. (2011).
\newblock On graph-based name disambiguation.
\newblock {\em Journal of Data and Information Quality (JDIQ)}, 2(2):10.

\bibitem[Fellegi and Sunter, 1969]{fellegi69}
Fellegi, I.~P. and Sunter, A.~B. (1969).
\newblock A theory for record linkage.
\newblock {\em Journal of the American Statistical Association}, 64:1183--1210.

\bibitem[Ferreira et~al., 2012]{ferreira2012brief}
Ferreira, A.~A., Gon{\c{c}}alves, M.~A., and Laender, A.~H. (2012).
\newblock A brief survey of automatic methods for author name disambiguation.
\newblock {\em Acm Sigmod Record}, 41(2):15--26.

\bibitem[Ferreira et~al., 2010]{ferreira2010effective}
Ferreira, A.~A., Veloso, A., Gon{\c{c}}alves, M.~A., and Laender, A.~H. (2010).
\newblock Effective self-training author name disambiguation in scholarly
  digital libraries.
\newblock In {\em Proceedings of the 10th annual joint conference on Digital
  libraries}, pages 39--48. ACM.

\bibitem[Friedman, 2001]{friedman2001greedy}
Friedman, J.~H. (2001).
\newblock Greedy function approximation: a gradient boosting machine.
\newblock {\em Annals of statistics}, pages 1189--1232.

\bibitem[Gentil-Beccot et~al., 2009]{gentil2009information}
Gentil-Beccot, A., Mele, S., Holtkamp, A., O'Connell, H.~B., and Brooks, T.~C.
  (2009).
\newblock Information resources in high-energy physics: Surveying the present
  landscape and charting the future course.
\newblock {\em Journal of the American Society for Information Science and
  Technology}, 60(1):150--160.

\bibitem[Guyon et~al., 2002]{guyon2002gene}
Guyon, I., Weston, J., Barnhill, S., and Vapnik, V. (2002).
\newblock Gene selection for cancer classification using support vector
  machines.
\newblock {\em Machine learning}, 46(1-3):389--422.

\bibitem[Han et~al., 2004]{han2004two}
Han, H., Giles, L., Zha, H., Li, C., and Tsioutsiouliklis, K. (2004).
\newblock Two supervised learning approaches for name disambiguation in author
  citations.
\newblock In {\em Digital Libraries, 2004. Proceedings of the 2004 Joint
  ACM/IEEE Conference on}, pages 296--305. IEEE.

\bibitem[Huang et~al., 2006]{huang2006efficient}
Huang, J., Ertekin, S., and Giles, C.~L. (2006).
\newblock Efficient name disambiguation for large-scale databases.
\newblock In {\em Knowledge Discovery in Databases: PKDD 2006}, pages 536--544.
  Springer.

\bibitem[Jones et~al., 01  ]{scipy}
Jones, E., Oliphant, T., Peterson, P., et~al. (2001--).
\newblock {SciPy}: Open source scientific tools for {Python}.
\newblock [Online; accessed 2015-08-10].

\bibitem[Kang et~al., 2009]{kang2009co}
Kang, I.-S., Na, S.-H., Lee, S., Jung, H., Kim, P., Sung, W.-K., and Lee, J.-H.
  (2009).
\newblock On co-authorship for author disambiguation.
\newblock {\em Information Processing \& Management}, 45(1):84--97.

\bibitem[Lange and Naumann, 2011]{lange2011frequency}
Lange, D. and Naumann, F. (2011).
\newblock Frequency-aware similarity measures: why arnold schwarzenegger is
  always a duplicate.
\newblock In {\em Proceedings of the 20th ACM international conference on
  Information and knowledge management}, pages 243--248. ACM.

\bibitem[Levin et~al., 2012]{levin2012citation}
Levin, M., Krawczyk, S., Bethard, S., and Jurafsky, D. (2012).
\newblock Citation-based bootstrapping for large-scale author disambiguation.
\newblock {\em Journal of the American Society for Information Science and
  Technology}, 63(5):1030--1047.

\bibitem[Liu et~al., 2014]{liu2014author}
Liu, W., Islamaj~Do{\u{g}}an, R., Kim, S., Comeau, D.~C., Kim, W., Yeganova,
  L., Lu, Z., and Wilbur, W.~J. (2014).
\newblock Author name disambiguation for pubmed.
\newblock {\em Journal of the Association for Information Science and
  Technology}, 65(4):765--781.

\bibitem[Louppe et~al., 2013]{louppe2013understanding}
Louppe, G., Wehenkel, L., Sutera, A., and Geurts, P. (2013).
\newblock Understanding variable importances in forests of randomized trees.
\newblock In {\em Advances in Neural Information Processing Systems}, pages
  431--439.

\bibitem[Malin, 2005]{malin2005unsupervised}
Malin, B. (2005).
\newblock Unsupervised name disambiguation via social network similarity.
\newblock In {\em Workshop on link analysis, counterterrorism, and security},
  volume 1401, pages 93--102.

\bibitem[McRae-Spencer and Shadbolt, 2006]{mcrae2006also}
McRae-Spencer, D.~M. and Shadbolt, N.~R. (2006).
\newblock Also by the same author: Aktiveauthor, a citation graph approach to
  name disambiguation.
\newblock In {\em Proceedings of the 6th ACM/IEEE-CS joint conference on
  Digital libraries}, pages 53--54. ACM.

\bibitem[Newman, 2001]{newman2001structure}
Newman, M.~E. (2001).
\newblock The structure of scientific collaboration networks.
\newblock {\em Proceedings of the National Academy of Sciences},
  98(2):404--409.

\bibitem[Pedregosa et~al., 2011]{scikitlearn}
Pedregosa, F., Varoquaux, G., Gramfort, A., Michel, V., Thirion, B., Grisel,
  O., Blondel, M., Prettenhofer, P., Weiss, R., Dubourg, V., Vanderplas, J.,
  Passos, A., Cournapeau, D., Brucher, M., Perrot, M., and Duchesnay, E.
  (2011).
\newblock Scikit-learn: Machine learning in {P}ython.
\newblock {\em Journal of Machine Learning Research}, 12:2825--2830.

\bibitem[Philips, 2000]{doublemetaphone}
Philips, L. (2000).
\newblock The double metaphone search algorithm.
\newblock {\em C/C++ Users J.}, 18(6):38--43.

\bibitem[Ruggles et~al., 2008]{rugglesintegrated}
Ruggles, S., Sobek, M., Fitch, C.~A., Hall, P.~K., and Ronnander, C. (2008).
\newblock {\em Integrated public use microdata series}.
\newblock Historical Census Projects, Department of History, University of
  Minnesota.

\bibitem[Schulz et~al., 2014]{schulz2014exploiting}
Schulz, C., Mazloumian, A., Petersen, A.~M., Penner, O., and Helbing, D.
  (2014).
\newblock Exploiting citation networks for large-scale author name
  disambiguation.
\newblock {\em EPJ Data Science}, 3(1):1--14.

\bibitem[Smalheiser and Torvik, 2009]{smalheiser2009author}
Smalheiser, N.~R. and Torvik, V.~I. (2009).
\newblock Author name disambiguation.
\newblock {\em Annual review of information science and technology},
  43(1):1--43.

\bibitem[Soler, 2007]{soler2007separating}
Soler, J. (2007).
\newblock Separating the articles of authors with the same name.
\newblock {\em Scientometrics}, 72(2):281--290.

\bibitem[Song et~al., 2007]{song2007efficient}
Song, Y., Huang, J., Councill, I.~G., Li, J., and Giles, C.~L. (2007).
\newblock Efficient topic-based unsupervised name disambiguation.
\newblock In {\em Proceedings of the 7th ACM/IEEE-CS joint conference on
  Digital libraries}, pages 342--351. ACM.

\bibitem[Strotmann and Zhao, 2012]{strotmann2012author}
Strotmann, A. and Zhao, D. (2012).
\newblock Author name disambiguation: What difference does it make in
  author-based citation analysis?
\newblock {\em Journal of the American Society for Information Science and
  Technology}, 63(9):1820--1833.

\bibitem[Taft, 1970]{nysiis}
Taft, R.~L. (1970).
\newblock Name search techniques.
\newblock Technical Report Special Report No. 1, New York State Identification
  and Intelligence System, Albany, NY.

\bibitem[{The National Archives}, 2007]{Soundex}
{The National Archives} (2007).
\newblock The soundex indexing system.

\bibitem[Torvik and Smalheiser, 2009]{torvik2009author}
Torvik, V.~I. and Smalheiser, N.~R. (2009).
\newblock Author name disambiguation in medline.
\newblock {\em ACM Transactions on Knowledge Discovery from Data (TKDD)},
  3(3):11.

\bibitem[Tran et~al., 2014]{tran2014author}
Tran, H.~N., Huynh, T., and Do, T. (2014).
\newblock Author name disambiguation by using deep neural network.
\newblock In {\em Intelligent information and database systems}, pages
  123--132. Springer.

\bibitem[Treeratpituk and Giles, 2009]{treeratpituk2009disambiguating}
Treeratpituk, P. and Giles, C.~L. (2009).
\newblock Disambiguating authors in academic publications using random forests.
\newblock In {\em Proceedings of the 9th ACM/IEEE-CS joint conference on
  Digital libraries}, pages 39--48. ACM.

\bibitem[Treeratpituk and Giles, 2012]{treeratpituk2012name}
Treeratpituk, P. and Giles, C.~L. (2012).
\newblock Name-ethnicity classification and ethnicity-sensitive name matching.
\newblock In {\em AAAI}. Citeseer.

\bibitem[Ward~Jr, 1963]{ward1963hierarchical}
Ward~Jr, J.~H. (1963).
\newblock Hierarchical grouping to optimize an objective function.
\newblock {\em Journal of the American statistical association},
  58(301):236--244.

\end{thebibliography}

\end{document}